\documentclass[aps,prl,twocolumn,showpacs,amsmath,amssymb,superscriptaddress]{revtex4}
\usepackage{epsfig}
\usepackage{bm}
\def\be{\begin{equation}}
\def\ee{\end{equation}}

\begin{document}
\title{Quantum-to-Classical Crossover in Full Counting Statistics}
\author{Eugene V. Sukhorukov}
\affiliation{D\'epartement de Physique Th\'eorique, Universit\'e de Gen\`eve,
CH-1211 Gen\`eve 4, Switzerland}
\author{Oleg M. Bulashenko}
\affiliation{Departament de F\'{\i}sica Fonamental,
Universitat de Barcelona, Diagonal 647, E-08028 Barcelona, Spain}

\date{\today}

\begin{abstract}
The reduction of quantum scattering leads to the suppression of shot noise. 
In the present paper, we analyze the crossover from the quantum transport regime 
with universal shot noise, to the classical regime where noise vanishes.
By making use of the stochastic path integral approach, we find the statistics of 
transport and the transmission properties of a chaotic cavity as a function of 
a system parameter controlling the crossover. We identify three different 
scenarios of the crossover.

\end{abstract}

\maketitle

Random transfer of charge in electrical conductors leads to time-dependent 
fluctuations of the current, a phenomenon called shot noise. 
In recent years, the shot noise has been extensively studied in mesoscopic conductors 
\cite{BB}, small degenerate electron systems of a size comparable to
the coherence length of electrons. In contrast to the classical 
shot noise in vacuum tubes, which was explained 
by Schottky already in 1918 \cite{SCH}, the shot noise in mesoscopic 
conductors originates from the quantum-mechanical scattering of electrons.  
Consequently, in a noninteracting mesoscopic conductor biased
by the chemical potential difference $\Delta\mu$, 
the average current $\langle I\rangle=
\Delta\mu\sum_n T_n$ \cite{buttiker}
(setting electron charge and the Planck constant $e=h=1$),
the noise power $S\equiv\langle\langle I^2\rangle\rangle 
=\Delta\mu\sum_n T_n(1-T_n)$ at zero temperature \cite{BB}, and in general
the higher cumulants of current $\langle\langle I^m\rangle\rangle$
\cite{LL1}, are determined by the transmission matrix $\hat t$, 
namely by the eigenvalues $T_n$, $n=1,\dots,N$, 
of $\hat t^{\dagger}\hat t$.
The current cumulants
$\langle\langle I^m\rangle\rangle=\Delta\mu N{\cal C}_m$
can be expressed via the cumulant generating function (CGF)
${\cal C}_m=\partial^m{\cal H}(\lambda)/\partial\lambda^m|_{\lambda=0}$.
In the semiclassical limit, $N\gg 1$, the CGF is given by
\cite{LL2}:
\be
{\cal H}(\lambda)=\int_0^1 dT\rho(T)\ln[1+T(e^{\lambda}-1)],
\label{CALH}
\ee  
where $\rho(T)=N^{-1}\sum_n\delta(T-T_n)$ is the transmission eigenvalue 
distribution. Eq.\ (\ref{CALH}) generalizes the binomial
statistics, and together with the inverse formula (\ref{rhogen}), 
provides a connection between the full counting statistics (FCS) 
and the scattering properties of a mesoscopic system to leading order 
in $1/N$.

The quantum origin of shot noise in mesoscopic conductors
implies that regardless of the character of disorder,
current can flow without noise if quantum scattering 
is suppressed  \cite{Houten}. 
Therefore, in the classical limit the noise should
vanish even in a chaotic system, such as a mesoscopic cavity, 
where the transport in quantum regime is universally described 
by Random Matrix Theory (RMT) \cite{RMT}. It has been predicted 
\cite{agam00} that in a mesoscopic cavity with long-range disorder 
that models chaotic dynamics
the noise power 
shows an exponential crossover $S=S_{\rm RMT}\exp(-\tau_E/\tau_D)$ 
as a function of the ratio of the Ehrenfest (diffraction) time 
$\tau_E$ to the average dwell time of electrons 
$\tau_D$. Ref.\ \cite{silvestrov} suggested that this crossover 
results from a sharp cut-off introduced by the Ehrenfest time 
in the exponential distribution ${\cal P}(t)=\tau_D^{-1}\exp(-t/\tau_D)$ 
of the dwell times of classical trajectories. 
Recent numerical analysis \cite{jacquod04} has demonstrated 
that the cut-off leads to a complete separation of the cavity's 
phase space into the quantum universal part of relative
volume $v=\exp(-\tau_E/\tau_D)$ and  the classical noiseless part 
of the volume $1-v$. As a result the eigenvalue distribution
splits in two terms, $\rho=v\rho_{\rm RMT}+(1-v)\rho_{\rm cl}$,
where 
\be
\rho_{\rm RMT}(T)=\frac{1}{\pi\sqrt{T(1-T)}}
\label{RMT}
\ee
is the universal RMT result, and $\rho_{\rm cl}(T)=[\delta(T)+\delta(1-T)]/2$
is the classical distribution. The onset of the quantum-to-classical 
crossover has been observed in the experiment on the shot noise of a 
mesoscopic cavity \cite{oberholzer02}. Since then interest 
in the physics 
of the crossover has grown dramatically and brought new results in the context 
of the shot noise suppression \cite{silvestrov, tworzydlo}, 
the proximity effect in Andreev billiards \cite{beenakker}, 
mesoscopic conductance fluctuations \cite{jacquod04,tworzydlo04}, and many other phenomena. 
 
In this Letter, we demonstrate that the presence of the homogeneous short-range disorder
in a chaotic cavity dramatically changes the quantum-to-classical 
crossover. It leads to the large-angle quantum scattering of electrons and results
in the relaxation of the deterministic occupation function $f_{\bf p}$, 
which takes values $0$ and $1$, to its fully quantum isotropic value $f_C<1$. 
The relaxation with the constant rate $\tau_Q^{-1}$, 
where $\tau_Q$ is the quantum scattering time, does not introduce a sharp cut-off 
in the dwell time distribution. As a result, all cumulants have a power-law 
dependence on the crossover parameter $\gamma=\tau_Q/\tau_D$. 
In particular, in contrast to the case of the long-range disorder discussed above,
the noise  power shows the power-law crossover \cite{sukhor}
\be
S=S_{\rm RMT}/(1+\gamma),\quad \gamma=\tau_Q/\tau_D.
\label{noise-power}
\ee
The distribution $\rho(T)$
gradually evolves as a function of the parameter $\gamma$ from its RMT limit (\ref{RMT})  
in the quantum regime to the classical limit $\rho_{\rm cl}$ with two
$\delta$-peaks \cite{footnote}.
\begin{figure}
\epsfxsize=7.0cm
\epsfbox{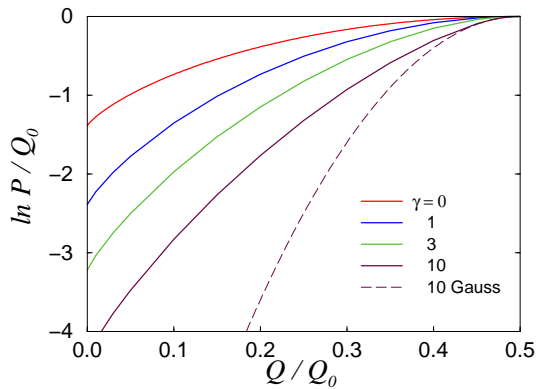}
\caption{The logarithm of the distribution of transmitted charge $Q$ 
plotted versus charge normalized to its maximum
value $Q_0=\Delta\mu Nt$. It is symmetric around the average value
$Q/Q_0=0.5$. Note a relatively weak dependence of the extreme
value statistics on the crossover parameter
$\gamma=\tau_Q/\tau_D$. The dashed line is the Gaussian distribution
shown for a comparison. 
}
\label{fig-cav} \end{figure}

{\em The model.}---The mesoscopic chaotic cavity is a metallic island
connected to the leads $L$ and $R$ through ballistic 
point contacts. 
This system has several characteristic time scales.
The time of the ballistic flight of electrons across the cavity
$\tau_F$ is much shorter than the average dwell time 
$\tau_D=n_F/N$, where $n_F$ is the density of states in the
cavity at Fermi level, and $N$ is the number of modes in each point contact
(symmetric cavity). 
We consider the quasi-ballistic regime, $\tau_Q\geq\tau_F$, where $\tau_Q$
is the time of {\em quantum} scattering off a short-range disorder, and neglect 
inelastic processes. The temperature is smaller than the bias and set here
to zero. 

In the semiclassical limit considered here, $N\gg 1$ and $\Delta\mu n_F\gg 1$, 
the transport to leading order in number of modes $N$
can be described classically. We use the classical approach of Refs.\ 
\cite{pilgram02,jordan04} based on the principle of minimal correlations \cite{blanter00}. 
According to this principle, the currents through the left and right point contacts 
$I_{L,R}$ are considered to be noise sources which are correlated solely via the 
conservation of the charge $\Delta\mu n_Ff_C$ in the cavity. Here 
$f_C=\langle f_{\bf p}\rangle$ is the isotropic part of the  
occupation function $f_{\bf p}$, and $\langle\ldots\rangle$ denotes the averaging 
over the momentum on the Fermi surface. The statistics of sources can be
obtained by taking into account the fermionic statistics of electrons, which leads to
binomial fluctuations of the occupation function in each semiclassical state 
with the cumulant generator $\ln[1+f_{\bf p}(e^\lambda-1)]$. 
Multiplying this function by the electron velocity $\bf v$, summing over ${\bf p}$, 
and integrating over the area of the contacts we obtain the generators 
$\Delta\mu N{\cal H}_{L,R}$ of the left and right current
\be
{\cal H}_l(\lambda_l,f_C)=\langle\ln[1+f_{\bf p}(e^{\lambda_l}-1)]
\rangle-f_l\lambda_l,\quad l=L,R,
\label{leads}
\ee
where $f_{L}=1$ and $f_R=0$ are the occupations in the left and right lead. This expression 
is the semiclassical limit of the result of Ref.\ \cite{LL1}. The charge conservation can be taken 
into account nonperturbatively in fluctuations $\delta f_C$ using the stochastic path integral 
\cite{pilgram02,jordan04}. In the stationary limit $t\gg \tau_D$, the saddle-point evaluation 
amounts to the minimization of the function 
%
%
\begin{equation}
{\cal H}={\cal H}_L(\lambda_C-\lambda/2,f_C)+
{\cal H}_R(\lambda_C+\lambda/2,f_C)
\label{function}
\end{equation}
with respect to the occupation $f_C$ and variable $\lambda_C$, 
a Lagrange multiplier conserving charge.
The result of this procedure gives the CGF (\ref{CALH}).

{\em Counting statistics.}---The crossover from the classical 
to quantum transport regime may be viewed as being caused by the relaxation of the 
classical occupation $f_{\bf p}=0,1$ to the quantum isotropic value $f_C$ 
as a result of scattering off the short-range disorder. This process
can be described by the Boltzmann equation
\begin{equation}
{\bf v} \nabla f_{\bf p} + \tau_Q^{-1}(f_{\bf p}-f_C) =0,
\label{Boltzmann}
\end{equation}
where the classical chaotic dynamics of electrons is taken into account by the
``gradient'' term and the quantum scattering is described by the second, 
the collision integral in the scattering time approximation.
In the classical limit $\gamma=\tau_Q/\tau_D\gg 1$ the second term can be neglected,
and the solution of Eq.\ (\ref{Boltzmann}) takes one of the boundary values
$f_{L,R}=0,1$ giving ${\cal H}_l=(f_C-f_l)\lambda_l$. Then the minimization
of the function (\ref{function}) leads to ${\cal H}=\lambda/2$ giving the average current
$\langle I\rangle=\Delta\mu N/2$ and no noise. 
In the quantum limit $\tau_Q/\tau_D\ll 1$ 
the second term in Eq.\ (\ref{Boltzmann}) dominates, 
therefore in Eq.\ (\ref{leads}) $f_{\bf p}$ may be replaced with $f_C$. 
Minimizing ${\cal H}$ given by (\ref{function}), we obtain the result \cite{pilgram02} 
\be
{\cal H}=2\ln(1+e^{\lambda/2})-2\ln2,\qquad \gamma=0,
\label{FCS1}
\ee
which agrees with the RMT result \cite{blanter01}. 

In order to obtain the corse-grained value of the logarithm 
in Eq.\ (\ref{leads}) we multiply Eq.\ (\ref{Boltzmann})
by $(f_{\bf p}-f_C)^{k-1}$ and integrate the resulting equation
\begin{equation}
\nabla [{\bf v} (f_{\bf p}-f_C)^k]
+(k/\tau_Q)(f_{\bf p}-f_C)^k=0,
\end{equation}
over the phase space $({\bf p},{\bf r})$ of the cavity.
Using the identity $\int d{\bf r}\nabla {\bf v}(\ldots)=
\int d{\bf s}{\bf v}\, (\ldots)$ 
we reduce the volume integral in the first term to the surface integral
over the cavity openings and arrive at the following expression
\be 
\langle (f_{\bf p}-f_C)^k \rangle 
=\frac{\gamma}{k+2\gamma}\,\left[(1-f_C)^k+(-f_C)^k\right].
\label{fk}
\ee
Expanding the logarithm in Eq.\ (\ref{leads}) in powers of $f_{\bf p}-f_C$,
using the result (\ref{fk}) and resumming the logarithm again, we obtain
the integral representation
\begin{eqnarray}
{\cal H}_l(\lambda_l) =&&
\hspace{-.3cm}-f_l\lambda_l+
\gamma\int_0^1 du \, u^{2\gamma-1} \nonumber\\ &&
\hspace{-.3cm}\times 
\left(
\ln\{1+[u+(1-u)f_C](e^{\lambda_l}-1)\}\right.\nonumber\\  
&&\quad \left.+\ln\{1+[(1-u)f_C](e^{\lambda_l}-1)\} 
\right).\,
\label{Hasym}
\end{eqnarray}
This expression has to be substituted into the variation function
(\ref{function}). Surprisingly, the stationary point 
is given by $\lambda_C=0$ and $f_C=1/2$ independent of $\gamma$, 
implying the absence of cascade corrections
\cite{nagaev02-1} to the FCS. 
Evaluating  Eq.\ (\ref{Hasym}) at the stationary point, 
we obtain the current generator for a symmetric cavity as a function of the 
crossover parameter $\gamma$:
\begin{equation} 
{\cal H}(\lambda,\gamma) = \lambda/2 
+ 2 \int_0^1 du \, \frac{u^{2\gamma+1} }{\coth^2(\lambda/4)-u^2}.
\label{Hsym}
\end{equation}
This equation is one of our main results. It correctly reproduces
the quantum limit (\ref{FCS1}) at $\gamma\to 0$ and has an 
asymptotic form ${\cal H}=\lambda/2+\gamma^{-1}\sinh^2 (\lambda/4) + O(\gamma^{-2})$
in the classical limit $\gamma\to\infty$ [valid for $\lambda\alt \ln(\gamma)$]. 
Thus the crossover has a power-law character
in contrast to the case of a long-range disorder: High cumulants  
are suppressed as $\sim \gamma^{-1}$ in the classical limit.

Normalized current cumulants ${\cal C}_m=\langle\langle I^m\rangle\rangle/(\Delta\mu N)$
may be obtained by differentiating the CGF (\ref{Hsym}) with respect to $\lambda$. 
Odd cumulants vanish ${\cal C}_m=0$ for $m\geq 3$ as a consequence of 
the zero temperature limit and of the fact 
that the cavity is symmetric. 
The first three non-vanishing cumulants are:
${\cal C}_1=1/2$, which determines the mean current, 
${\cal C}_2=1/[8(\gamma+1)]$, which determines the noise 
power and agrees with the result (\ref{noise-power}), 
and ${\cal C}_4=(\gamma-1)/[32(\gamma+1)(\gamma+2)]$.

The logarithm of the distribution of transmitted charge 
in the stationary phase approximation is given by 
$\ln P(Q)=Q_0\,{\rm min}_\lambda\{{\cal H}(\lambda)-{\cal Q}
\lambda\}$ \cite{jordan04},
where ${\cal Q}=Q/Q_0$ is the transmitted charge normalized
to its maximum value $Q_0=\Delta\mu Nt$.
The result of the evaluation using Eq.\ (\ref{Hsym}) is shown 
in Fig.\ \ref{fig-cav}
for different values of $\gamma$.
In the quantum limit, we use Eq.\ (\ref{FCS1}) 
to obtain $\ln P(Q)/Q_0=-2\ln 2-2[{\cal Q}\ln {\cal Q}
+(1-{\cal Q})\ln(1-{\cal Q})]$, which vanishes at the 
average value of charge ${\cal Q}=1/2$, giving the correct 
normalization of $P(Q)$. In the classical limit $\gamma\gg 1$
the noise is Gaussian, $\ln P(Q)/Q_0=-4\gamma\,({\cal Q}-1/2)^2$,
for $|{\cal Q}-1/2|\alt\gamma^{-1}$.
Surprisingly, the extreme value distribution 
in the range $\gamma^{-1}\alt |{\cal Q}-1/2|\alt 1/2$
shows a weak $\gamma$ dependence:
$\ln P(Q)/Q_0=-2|{\cal Q}-1/2|
\{\ln(8\gamma|{\cal Q}-1/2|)-1\}$,
see Fig.\ \ref{fig-cav}.
This remarkable behavior may be attributed to the formation
of almost open (closed) quantum channels, the situation 
specific to the short-range disorder considered here. 
The number of such channels is 
nearly independent of $\gamma$ and close to the total 
number of modes $N$ (see the discussion below).
\begin{figure}
\epsfxsize=7.0cm
\epsfbox{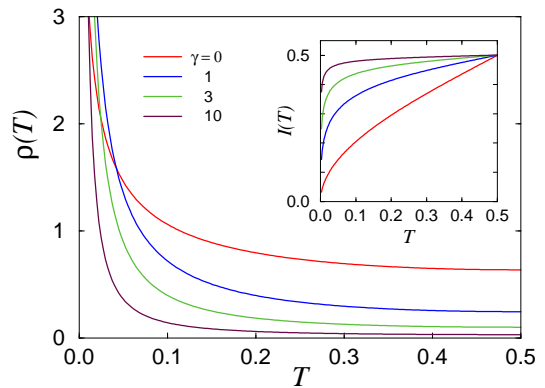}
\caption{The crossover of the distribution of transmission eigenvalues $\rho(T)$
between the quantum ($\gamma$=0) and classical ($\gamma\to\infty$) regimes.
$\rho(T)$ is symmetric around $T$=0.5. 
Inset: Integrated probability distribution ${\cal I}(T)=\int_0^T \rho(T')dT'$
for the same set of parameters.}
\label{fig-rho} \end{figure}

{\em Distribution of transmission eigenvalues.}---Having found 
the CGF, we now invert Eq.\ (\ref{CALH}) 
in order to obtain the distribution of transmission eigenvalues 
$\rho(T)$. We note that ${\cal H}$ as a function of 
the variable $\Lambda\equiv e^{\lambda}-1$ has a branch cut 
in the complex plane at $-\infty<\Lambda<-1$. 
Analytically continuing 
from $\Lambda>1$ to the branch cut \cite{Nazarov}, 
we obtain:
\begin{equation} 
\rho(T) = \frac{1}{\pi T^2} \,
{\rm Im} \left( \left.
\partial{\cal H}/\partial\Lambda
\right|_{\Lambda\to -1/T-i0}\right).
\label{rhogen}
\end{equation}
Using this relation together with Eq.\ (\ref{Hsym}),
we arrive at the following result for 
$\rho(T)$ in the crossover regime:
\be 
\rho(T) = \frac{\gamma}{\pi\sqrt{T(1-T)}}  
\int_{-1}^1 du\, \frac{(1-u^2)\,|u|^{2\gamma-1}} {(1+u)^2-4Tu}.
\label{rhosym}
\ee
The distribution is symmetric with respect to 
$T\to 1-T$ and properly
normalized: one can verify that $\int_0^1 dT\rho(T)=1$.
In the quantum limit $\gamma\to 0$ Eq.\ (\ref{rhosym}) leads to the
RMT result (\ref{RMT}).
In the classical limit we obtain the asymptotic formula
\begin{equation} 
\rho(T)|_{\gamma\to\infty} = 
\frac{1}{8\pi\gamma\,[T(1-T)]^{3/2}} + O(\gamma^{-2}),
\label{rhosyminf}
\end{equation}
which is valid away from the points $T$=0 and $T$=1 where $\rho(T)$ in
(\ref{rhosyminf}) is 
divergent. This divergence, being cut at $T,1-T\sim \gamma^{-2}$ in Eq.\ (\ref{rhosym}), 
gives the main contribution to the normalization of $\rho(T)$, to the average current,
and determines the extreme value statistics discussed above. 
However, it is integrable for high cumulants of current.

The distribution $\rho(T)$ for several values of $\gamma$
is illustrated in Fig.\ \ref{fig-rho}. The quantum-to-classical crossover 
appears as a gradual transition from the RMT distribution at $\gamma=0$ to 
two $\delta$ functions at $T=0$ and $T=1$. Following Ref.\ \cite{jacquod04}
we plot the integrated distribution ${\cal I}(T)=\int_0^T \rho(T')dT'$,
which turns out to be a smooth function of $T$. This is in contrast to the case of 
a long-range disorder, where ${\cal I}(T)$ shows an offset at $T=0$ \cite{jacquod04}, 
indicating the separation of phase space into a classical and a quantum part.   
Our result implies that such a separation does not occur in the case of a 
homogeneous short-range disorder.

{\em Inhomogeneous disorder.}---So far we have considered  a relatively
weak homogeneous disorder with the strength characterized by the 
scattering time $\tau_Q$. 
Another experimentally relevant situation is the case of a strong
inhomogeneous short-range disorder.
For instance, a few strong impurities, sharp openings to the leads 
or irregularities at the boundary 
of the cavity belong to this class of disorder. 
The inhomogeneity implies that some trajectories 
do not enter the disordered region and remain classical with $f_{\bf p}=0,1$. 
The fact that the disorder is 
strong means that all trajectories entering the disordered region 
acquire the isotropic occupation $f_C$. For the corse-grained 
occupation function $\langle f_{\bf p}\rangle$ this leads to the relaxation 
with the collision rate $\tau_{\rm imp}$. This process is described by Eq.\ 
(\ref{Boltzmann})
with $\tau_Q$ replaced by $\tau_{\rm imp}$. 
In the present case the solution of this equation determines the relative volume of 
the quantum phase space $v=1/(1+\gamma)$, where $\gamma=\tau_{\rm imp}/\tau_D$
is the new crossover parameter. Therefore we conclude that 
inhomogeneous strong disorder leads to the complete separation of the phase space 
on the classical and quantum parts with the consequence that 
$\rho=v\rho_{\rm RMT}+(1-v)\rho_{\rm cl}$. However, in contrast to the case of 
long-range disorder, the FCS is a power-law function of the crossover parameter: 
${\cal C}_n(\gamma)={\cal C}_n(0)/(1+\gamma)$.         

{\em Asymmetric cavity.}---From the above analysis it follows that 
the noise power has the same dependence (\ref{noise-power}) 
on the crossover parameter $\gamma$ for both types of a short-range disorder.
Moreover, since the odd cumulants of current for the symmetric cavity vanish at zero 
temperature, the difference in noise appears starting from 
the fourth cumulant. The recent progress in the measurement of a third-order cumulant 
\cite{reulet03} motivates us to analyze the counting statistics for 
asymmetric cavities with non-equal number of modes $N_L$ and $N_R$ in
the point contacts, for which odd cumulants are expected to be finite \cite{b04}. 

To obtain the third cumulant for the case of a homogeneous disorder 
we utilize the operator approach of Ref.\ \cite{jordan04}, 
which in the case of a single variable represents a convenient alternative to the 
cascade diagrammatics [\onlinecite{nagaev02-1},\onlinecite{jordan04}]. Omitting lengthy calculations
we present the result 
\be
{\cal C}_3(\gamma)=\frac{3{\cal C}_3(0)}{(1+\gamma)(3+2\gamma)},
\label{third-cum} 
\ee
where ${\cal C}_3(0)=-2[N_LN_R(N_L-N_R)/(N_L+N_R)^3]^2$ is the RMT value of the third cumulant.
We note that the third cumulant (\ref{third-cum}) vanishes as 
$\gamma^{-2}$ in the classical limit, i.e.\ faster than the one for the 
inhomogeneous disorder. 
Therefore the measurement of the third cumulant may help to distinguish the 
character of disorder.

In conclusion, we have analysed the FCS and transmission properties 
of a mesoscopic cavity at the crossover from the universal quantum to 
the classical transport regimes.
We have found new different scenarios of the crossover in a cavity with short-range
disorder. In case of homogeneous disorder, the crossover occurs via the formation of almost
open (closed) quantum channels, which determine the extreme value statistics. In the case 
of an inhomogeneous strong disorder, the phase space of the cavity splits in two parts: 
classical noiseless channels and quantum RMT channels. In both cases the FCS has a 
power-law dependence on the crossover parameter. 

We would like to thank Ph.\ Jacquod for helpful discussions. 
This work was supported by the SNF, and by INTAS 
(project 0014, open call 2001). O.M.B.\ acknowledges support by 
the MCyT of Spain through the ``Ram\'on y Cajal'' program.


\begin{thebibliography}{}

\bibitem{BB}
For a review, see Ya.M. Blanter and M. B\"uttiker, Phys. Rep. {\bf 336}, 1 (2000);
C. Beenakker and C. Sch\"onenberger, Phys. Today {\bf 56}, 37 (2003).

\bibitem{SCH}
W. Schottky, Ann. Phys. (Leipzig) {\bf 57} 541 (1918).

\bibitem{buttiker} M. B\"uttiker, IBM J. Res. Dev. {\bf 32}, 317 (1988).

\bibitem{LL1} 
L.S. Levitov and G.B. Lesovik, JETP Lett. {\bf 58}, 230 (1993).

\bibitem{LL2}
H. Lee, L.S. Levitov, and A.Yu. Yakovets, Phys. Rev. B {\bf 51}, 4079 (1995). 

\bibitem{Houten} C. W. J. Beenakker and H. van Houten, Phys. Rev. B
{\bf 43} 12066 (1991). 

\bibitem{RMT}
C. W. J. Beenakker,
Rev. Mod. Phys. {\bf 69}, 731 (1997).

\bibitem{agam00} 
O. Agam, I. Aleiner, and A. Larkin,
Phys. Rev. Lett. {\bf 85}, 3153 (2000).

\bibitem{silvestrov} 
P.G. Silvestrov, M.C. Goorden, and C.W.J. Beenakker,
Phys. Rev. B {\bf 67}, 241301 (2003).

\bibitem{jacquod04} 
Ph. Jacquod and E.V. Sukhorukov,
Phys. Rev. Lett. {\bf 92}, 116801 (2004).

\bibitem{oberholzer02}
S. Oberholzer, E.V. Sukhorukov, and C. Sch\"onenberger,
Nature {\bf 415}, 765 (2002).

\bibitem{tworzydlo} 
J. Tworzyd{\l}o, A. Tajic, H. Schomerus, and 
C.W.J. Beenakker, Phys. Rev. B {\bf 68}, 115313 (2003).

\bibitem{beenakker} For a review, see
C.W.J. Beenakker, cond-mat/0406018.

\bibitem{tworzydlo04} 
J. Tworzydlo, A. Tajic, and C.W.J. Beenakker, 
Phys. Rev. B {\bf 69}, 165318 (2004).

\bibitem{sukhor}
E. Sukhorukov, in {\em Fluctuation phenomena in low dimensional conductors},
S. Oberholzer, Ph.D. thesis, Appendix B (University of Basel, 2001).

\bibitem{footnote}
Here we assume $\tau_E\gg\tau_Q$. Combined effect of short-range
and long-range disorder will be addressed elsewhere.

\bibitem{pilgram02} 
S. Pilgram, A.N. Jordan, E.V. Sukhorukov, and M. B\"utti\-ker,
Phys. Rev. Lett. {\bf 90}, 206801 (2003).

\bibitem{jordan04} 
A.N. Jordan, E.V. Sukhorukov, and S. Pilgram, 
cond-mat/0401650.

\bibitem{blanter00} 
Ya.M. Blanter, E.V. Sukhorukov, Phys. Rev. Lett. {\bf 84}, 1280 (2000).

\bibitem{blanter01} 
Ya.M. Blanter, H. Schomerus, and C.W.J. Beenakker, 
Physica E {\bf 11}, 1 (2001).

\bibitem{nagaev02-1} 
K.E. Nagaev, Phys. Rev. B {\bf 66}, 075334 (2002);
K.E. Nagaev, P. Samuelsson, and S. Pilgram, 
Phys. Rev. B {\bf 66}, 195318 (2002).

\bibitem{Nazarov}
Yu. V. Nazarov, Phys. Rev. Lett. {\bf 73}, 134 (1994).

\bibitem{reulet03} 
B. Reulet, J. Senzier, and D.E. Prober,
Phys. Rev. Lett. {\bf 91}, 196601 (2003).

\bibitem{b04} 
O.M. Bulashenko, cond-mat/0403388.

\end{thebibliography}
\end{document}